\documentclass[submission]{eptcs}
\providecommand{\event}{PXTP 2015}
\usepackage{amsfonts}
\usepackage{url}

\title{The Common HOL Platform}
\author{Mark Adams
\institute{Proof Technologies Ltd, UK}
\institute{Radboud University, Nijmegen, The Netherlands}}
\def\titlerunning{The Common HOL API}
\def\authorrunning{M. Adams}

\def\star{\textsuperscript{*}}
\def\plus{\textsuperscript{+}}

\begin{document}
\maketitle

\begin{abstract}
The Common HOL project aims to facilitate porting source code and
proofs between members of the HOL family of theorem provers.  At the
heart of the project is the Common HOL Platform, which defines a
standard HOL theory and API that aims to be compatible with all HOL
systems.  So far, HOL Light and hol90 have been adapted for
conformance, and HOL Zero was originally developed to conform.
In this paper we provide motivation for a platform, give an overview
of the Common HOL Platform's theory and API components, and show how
to adapt legacy systems.  We also report on the platform's successful
application in the hand-translation of a few thousand lines of source
code from HOL Light to HOL Zero.
\end{abstract}

\section{Introduction}
\label{sec:introduction}

The HOL family of theorem provers started in the 1980s with
HOL88~\cite{hol}, and has since grown to include many systems, most
prominently HOL4~\cite{hol4}, HOL Light~\cite{hollight}, ProofPower
HOL~\cite{pp} and Isabelle/HOL~\cite{isabellehol}.  These four main
systems have developed their own advanced proof facilities and
extensive theory libraries, and have been successfully employed in
major projects in the verification of critical hardware and
software~\cite{clawz,sel4} and the formalisation of
mathematics~\cite{flyspeck}.

It would clearly be of benefit if these systems could ``talk'' to
each other, specifically if theory, proofs and source code could be
exchanged in a relatively seamless manner.  This would reduce the
considerable duplication of effort otherwise required for one system
to benefit from the major projects and advanced capabilities
developed on another.  Work to date has concentrated on exchange of
proofs via proof objects, with some degree of success, but little has
been done to facilitate porting of source code.

The Common HOL Platform is part of the Common HOL project for
facilitating the porting of source code and proofs between HOL systems.
It defines a standard HOL theory compatible with the core theory of each
HOL system, and an application programming interface (API) of
programming components that is more-or-less common to all HOL systems.
It has so far been supported in HOL Light, hol90~\cite{hol90} and HOL
Zero~\cite{holzeroweb}.

In this paper we give an overview of the platform.  In Section~2, we
further discuss motivation.  In Section~3, we cover the platform's
choice of components.  In Section~4, we explain how to adapt legacy
systems to conform to the platform.  In Section~5, we report on its
successful usage in assisting the manual porting of both new and
legacy source code.  In Section~6, we present our conclusions.

\section{Motivation}
\label{sec:motivation}

By definition, all systems in the HOL family implement the HOL logic
or a close variant.  However, in practice their commonality stretches
far beyond this.  They have broadly similar axiomatisations of the
logic, similar mechanisms for logical extension, similar formal
language concrete syntax and build up similar foundational theory.
Furthermore, in most basic usage at least, they each support similar
paradigms of user interaction, namely simple forwards-style
application of inference rules and backwards-style tactic proofs via
the subgoal goal package~\cite{subgoal}, performed in an interactive
functional programming session.  Also, their implementations are all
written in variants of the ML functional programming language, all
employ an LCF-style architecture~\cite{lcf} and are all built up from
similar libraries of programming utilities, syntax utilities,
inference rules and tactics.

Other than in these basic aspects, the systems branch off in their own
respects.  Each builds up considerable theory beyond the basic
foundations in its own way.  For example, real numbers in HOL Light
are constructed quite differently from real numbers in ProofPower HOL.
There is also much variation in their provision of user proof commands,
especially for those relating to proof automation, with each
system having its own strengths and idiosyncrasies.  Most different is
Isabelle/HOL, which is implemented as an instantiation of the
Isabelle generic theorem prover~\cite{isabelle} rather than by having
its deductive system ``hardwired'' as source code, and supports a
variant of the HOL logic that has axiomatic type classes.  Also, the
predominant mode of interaction with Isabelle has become the
declarative proof language Isar in conjunction with a bespoke IDE,
rather than the subgoal package in an interactive ML session.

Porting proofs between HOL systems by hand involves translating proofs 
scripts.  These proof scripts typically involve heavy use of high-level
proof commands that differ between systems.  In cases where such
commands are used to finish off subgoals, it is often possible to find
a suitably powerful command to do the same in the target system, but in
other cases proof scripts have to be recreated from scratch.  Automatic
proof porting, via recording of low-level proof steps and export to
proof object files, is vastly preferable if it can be made sufficiently
reliable.  Such a capability requires a platform of common foundational
theory, inference rules and logical extension mechanisms in both
systems.

There have been notable successes in the large scale porting of legacy
proofs between HOL systems via proof objects.  Obua and
Skalberg~\cite{obuaskalberg} developed a capability for porting proofs
from HOL4 to Isabelle/HOL, using a theory platform based on the HOL4
inference kernel, and then adapted this for porting from HOL Light to
Isabelle/HOL.  Kaliszyk and Krauss~\cite{cezary} developed a capability
for porting from HOL Light to Isabelle/HOL, based on the HOL Light
inference kernel.  The OpenTheory project~\cite{opentheory} is based
around the HOL Light axiomatisation, and establishes a common proof
object format for porting proofs between various HOL systems, including
HOL4, ProofPower/HOL and HOL Light, with ongoing work to support
Isabelle/HOL.  However, these capabilities would all struggle to port
something as large as the entire Flyspeck project~\cite{flyspeck}.
We believe that significant advances in capability can be achieved
by exploiting a broader commonality that exists between HOL systems,
using a platform at a somewhat higher level than the inference kernel
of one system.

Porting source code from one system to another currently requires
deep knowledge of both systems' implementations and can entail weeks
of effort to replicate behaviour sufficiently closely.  Naive porting
of high-level routines will typically result in unreliable code due
to the compounding of small and subtle differences in the theory or
in ML function behaviour.  We know of no pre-existing capability for
supporting the systematic porting of source code between HOL systems.

We believe that if the existing HOL systems can be adapted to support
a well-designed API that reflects the commonality of ``primary
functionality'' (by which we mean functionality directly concerned with
theorem proving) between the systems, then much of the pain of porting
source code can be avoided.  There is then a platform of precisely
corresponding programming components, and source code built on this
platform in one system can be trivially but accurately ported to
another system conforming to the same platform.  As is also the case
for a proof porting capability, both ML components and foundational
theory have to be taken into account when designing an effective
platform.

%

\section{Components}

In this section, we give an overview of the components that make
up version 0.5 of the Common HOL Platform.  This is the latest
version, and has been implemented for HOL Light and HOL Zero.  An
earlier version was implemented for hol90, but this has not yet been
upgraded.  Even though the platform has not yet been implemented for
ProofPower HOL or HOL4, it has been carefully designed with knowledge
of how these systems work.  However, little consideration has so far
been given to Isabelle/HOL, which presents greater challenges due to
its greater differences.  A significant redesign of the standard
would probably be required to properly cater for Isabelle/HOL.

There is no space in this paper to list all the platform components,
let alone to describe each one.  Instead we provide various tables
comparing some corresponding components from hol90, HOL4, ProofPower HOL,
HOL Light and HOL Zero.  For a given system, each platform
component is either exactly represented in the system, or it is
approximately represented, or it is not represented in the system.
In our listings, those components only approximately corresponding are
written in curly brackets.

There is not yet a single stand-alone document specifically for the
purpose of precisely defining each platform component.  However, part
of the original motivation for the HOL Zero system was to act as a
clear demonstration of the platform, and it has been designed to
exactly conform to platform behaviour without adaption.  Readers can
download the HOL Zero source distribution~\cite{holzeroweb}, where
source code file {\tt commonhol.mli} gives a complete list of the API
components, and the user manual appendices give a precise description
of each API and theory component.

\subsection{Considerations}
\label{sec:reqs}

Here we discuss some factors that should be taken into consideration
when choosing the components.

\begin{description}

  \item [Commonality]  Platform components should broadly reflect the
     commonality that exists between the systems.  Including
     components that are only relevant in one system would entail
     extra effort to make the other systems conformant, and would
     be of little use to them.  Not including components that are
     common to all systems would mean that basic components from one
     system would have to be needlessly considered when porting to a
     target system.

  \item [Usage]  Amount of usage in post-platform code should be
     taken into consideration when deciding the platform components.
     Heavily used components should almost qualify by default.

  \item [Level]  The components should be sufficiently high-level to
     be of likely use in post-platform source code.  For example,
     including low-level subcomponents used to make a HOL term
     parser would be of little use, even if these components were
     common to all HOL systems.

  \item [Precision]  A platform without precisely defined components of
     course loses much of its purpose.  In HOL systems, there are
     many small differences in the details of the behaviour of various
     corresponding basic functions.  For each component, the platform
     should explicitly specify its exact behaviour or otherwise be
     clear about what is not specified.  Non-conformant components must
     have platform-conformant variants defined as part of platform
     qualification.

  \item [Underspecification]  The API should allow some degree of
     flexibility in certain kinds of details about it components.
     For example, the ML names of the components, or the order in which
     function components take arguments and whether tuples or curried
     form is used.  The API should seek to minimise the effort
     required to make legacy systems conformant
     by underspecifying these details, which are not the kinds of
     differences that make porting source code difficult.

  \item [Completeness]  The components should be complete in the
     sense that all primary functionality can be built from
     platform components alone.  This becomes essential for the
     constructors and destructors of abstract datatypes (such as for
     HOL types, terms and theorems) because there is otherwise no way
     of manipulating such values.

  \item [Coherence]  The components should be chosen as a coherent
     set that categorise in a complete and consistent way and that
     composes robustly.  This makes it easier to write new code based
     on the API, as well as helping portability.

  \item [Performance]  The API should not exclude components that
     are important to the performance of a system if this means they
     would otherwise need to be reimplemented in the outer platform in
     terms of API components to result in a significant degradation
     in performance.
     
  \item [Ease of Implementation]  The implementation effort required
     to conform to a platform is a significant consideration.
     Otherwise, in practice the platform will not get implemented for
     the full range of HOL systems, which defeats its purpose.

\end{description}

\subsection{Theory Components}
\label{sec:theory}

The theory components are the axioms, declarations and definitions
that must exist in a conformant system's theory.  They must form a
sufficient basis for building up each HOL system's theory.

There is some variation in the systems' axiomatisations, especially
between HOL Light and the other systems.  Because each system
implements the same formal logic, for our purposes of completeness it
is sufficient to choose the core theory (i.e.\ the theory of
the logical core) of one system as the theory platform, and to derive
this in the other systems from their respective core theories.  The
outer platform (see Section~\ref{sec:arch}) in these other systems 
can then ``re-derive'' the system's core theory using the theory
platform.  A platform theorem may be an axiom or definition theorem
in one system and a derived theorem in another, but as far as the
platform is concerned they are all just theorems.

Our theory platform features the axioms and definitions of ProofPower
HOL, which we view as the most intuitive, and which are close to those
of hol90 and HOL4.  It also includes the HOL Light definition of the
implication operator, which does not feature in the other systems
because the behaviour of implication drops out from their primitive
inference rules and the implication antisymmetry axiom.  Including this
definition means that any of the systems' primitive inference rule set
suffices to complete the deductive system.  A handful of fundamental
theorems that are common to but derived in each system are included in
the platform, such as the truth theorem and the Law of the Excluded
Middle, because they are inevitably needed in implementing the platform
and so may as well feature as components.

The type constants and constants declared in the theory platform
include those from the basic theory about predicate logic and lambda
calculus that is common to each HOL system, established in the logical
core and initial derived theory of each system.  This includes the
function space type operator and the boolean base type, plus the
equality, conjunction, disjunction, implication and logical negation
operators, the universal, existential and unique existential
quantifiers and the Hilbert choice operator.

Beyond this, each system builds up essentially equivalent theory of
pairs, lists and natural numbers.  To take advantage of this
commonality, the platform also includes theory for pairs and natural
numbers, including natural number numerals and 13 classic arithmetic
operators including plus, multiply and exponentiation.  Theory for
lists does not currently feature, but is planned for inclusion in a
future version.

The representation of natural number numerals varies between HOL
systems: in HOL Light, HOL4 and HOL Zero, each numeral is constructed
using compounding of two unary operators on the zero constant (one for
multiplying by two and adding one, and one for multiplying by two and
adding zero or two depending on the system), whereas numerals in hol90
and ProofPower HOL form an infinite family of constants.  However,
beyond the definition of a set of basic numeral arithmetic evaluation
inference rules, these differences do not surface in practice in the
implementations of the systems.  Thus we have abstracted away from the
theory platform the detail of how numerals are defined.

\begin{table}
\footnotesize
\centering
\begin{tabular}{ c | c | c | c | c }
 hol90         & HOL4          & ProofPower    & HOL Light     & HOL Zero  \\
 \hline
 {\tt "bool"}  & {\tt "bool"}  & {\tt "BOOL"}  & {\tt "bool"}  & {\tt "bool"} \\
 {\tt "fun"}   & {\tt "fun"}   & {\tt "$\rightarrow$"}  & {\tt "fun"} & {\tt "->"} \\
 {\tt "prod"}  & {\tt "prod"}  & {\tt "$\times$"} & {\tt "prod"} & {\tt "\#"} \\
 {\tt "ind"}   & {\tt "ind"}   & {\tt "IND"}   & {\tt "ind"}   & {\tt "ind"} \\
 {\tt "num"}   & {\tt "num"}   & {\tt "$\mathbb{N}$"}  & {\tt "num"}    & {\tt "nat"} \\
 \hline
 {\tt "T"}     & {\tt "T"}     & {\tt "T"}     & {\tt "T"}     & {\tt "true"} \\
 {\tt "F"}     & {\tt "F"}     & {\tt "F"}     & {\tt "F"}     & {\tt "false"} \\
 {\tt "="}     & {\tt "="}     & {\tt "="}     & {\tt "="}     & {\tt "="} \\
 {\tt "/\char`\\"} & {\tt "/\char`\\"}
                               & {\tt "$\wedge$"} & {\tt "/\char`\\"} & {\tt "/\char`\\"} \\
 {\tt "\char`\\/"} & {\tt "\char`\\/"}
                               & {\tt "$\vee$"} & {\tt "\char`\\/"} & {\tt "\char`\\/"} \\
 {\tt "$\sim$"} & {\tt "$\sim$"} & {\tt "$\neg$"}  & {\tt "$\sim$"} & {\tt "$\sim$"} \\
 {\tt "!"}     & {\tt "!"}    & {\tt "$\forall$"} & {\tt "!"}  & {\tt "!"} \\
 {\tt "?"}     & {\tt "?"}    & {\tt "$\exists$"} & {\tt "?"}  & {\tt "?"} \\
 {\tt "?!"}    & {\tt "?!"}   & {\tt "$\exists_1$"} & {\tt "?!"} & {\tt "?!"} \\
 {\tt "@"}     & {\tt "@"}    & {\tt "$\varepsilon$"} & {\tt "@"}     & {\tt "@"} \\
 \hline
 {\tt IMP\_ANTISYM\_AX} & {\tt IMP\_ANTISYM\_AX\plus} & {\tt $\Rightarrow$\_antisym\_axiom}
                                                                             & -                      & {\tt imp\_antisym\_ax} \\
 {\tt ETA\_AX}          & {\tt ETA\_AX}          & {\tt $\eta$\_axiom}        & {\tt ETA\_AX}          & {\tt eta\_ax} \\
 {\tt SELECT\_AX}       & {\tt SELECT\_AX}       & {\tt $\varepsilon$\_axiom} & {\tt SELECT\_AX}       & {\tt select\_ax} \\
 {\tt BOOL\_CASES\_AX}  & {\tt BOOL\_CASES\_AX}  & {\tt bool\_cases\_axiom}   & {\tt BOOL\_CASES\_AX\plus}                                                                                                    & {\tt bool\_cases\_thm\plus} \\
 {\tt INFINITY\_AX}     & {\tt INFINITY\_AX}     & {\tt infinity\_axiom}      & {\tt INFINITY\_AX}     & {\tt infinity\_ax} \\
 {\tt T\_DEF}           & {\tt T\_DEF}           & {\tt t\_def}               & {\tt T\_DEF}           & {\tt true\_def} \\
 {\tt F\_DEF}           & {\tt F\_DEF}           & {\tt f\_def}               & {\tt F\_DEF}           & {\tt false\_def} \\
 {\tt AND\_DEF}         & {\tt AND\_DEF}         & {\tt $\wedge$\_def}        & {\tt \{AND\_DEF\}}     & {\tt conj\_def} \\
 -                      & -                      & -                          & {\tt IMP\_DEF}         & - \\
 {\tt OR\_DEF}          & {\tt OR\_DEF}          & {\tt $\vee$\_def}          & {\tt OR\_DEF}          & {\tt disj\_def} \\
 {\tt NOT\_DEF}         & {\tt NOT\_DEF}         & {\tt $\neg$\_def}          & {\tt NOT\_DEF}         & {\tt not\_def} \\
 {\tt FORALL\_DEF}      & {\tt FORALL\_DEF}      & {\tt $\forall$\_def}       & {\tt FORALL\_DEF}      & {\tt forall\_def} \\
 {\tt EXISTS\_DEF}      & {\tt EXISTS\_DEF}      & {\tt $\exists$\_def}       & {\tt EXISTS\_THM\plus} & {\tt exists\_def} \\
 {\tt \{UEXISTS\_DEF\}} & {\tt \{UEXISTS\_DEF\}} & {\tt $\exists_1$\_def}     & {\tt \{UEXISTS\_DEF\}} & {\tt uexists\_def} \\
\end{tabular}
\caption{The type constants, some of the constants and some of the
         theorems (including all the axioms) of the theory platform.
         Derived theorems in a given system are marked with
	 {\tt \footnotesize \plus}.}
\label{tab:thy3}
\end{table}

%
%

\subsection{API Components}
\label{sec:api}

The API components form the ML interface for programming primary
functionality.  There are approximately 475 components, mainly
consisting of ML function and constant values, but also seven datatypes
and three exceptions.  Three configuration values are also provided,
that hold the HOL system name and version and the Common HOL Platform
version.  In each conformant system, the API is provided as an ML module
interface file, with components given the same ordering to aid
comparison between systems.

Note that table components that have ML infix fixity in a given system
are written in parentheses.
%

\subsubsection{Functional Programming Library}
\label{sec:lib}

There are around 100 functional programming library components
(see Table~\ref{tab:fplib} for a selection).

Included are many basic operations on ML pairs, lists and strings,
such as selecting the first element of a pair, reversing the order of
elements in a list, or turning an integer into a string.  Association
lists are also supported.  Also included are various classic functional
programming meta operations, e.g.\ for applying a function to each
element in a set, or folding up a list into a single element by
repeated application of a binary operator.  There is also a collection
of set operations on lists, such as set membership and set union,
under either equality comparison or a supplied equivalence relation.

For coherence, we fill out the gaps that exist in the various legacy
systems' libraries.  For example, all kinds of folding operators and
their inverses, unfolding operators, are provided, and all set
operations are provided for both under equality and a supplied
equivalence relation.

Three kinds of standard exception are catered for: normal failure,
catastrophic failure and ``local failure'' (used for control flow
within a function).  The API underspecifies the form of the
exception arguments and the textual content of error messages

Note that there is some variation in the behaviour of some library
functions between systems.  For example, {\tt funpow}, which iterates
a function application for the number of times specified by a
supplied integer, does not fail in hol90, HOL4 or HOL Light if the
integer is negative.  Generally, platform functions are specified to
fail if supplied with invalid arguments, and the platform version of
{\tt funpow} fails if its supplied integer is negative, as is done in
ProofPower HOL and HOL Zero.

\begin{table}
\footnotesize
\centering
\begin{tabular}{ c | c | c | c | c }
 hol90         & HOL4          & ProofPower    & HOL Light     & HOL Zero  \\
 \hline
 {\tt curry}   & {\tt curry}   & {\tt curry}   & {\tt curry}   & {\tt curry} \\	
 {\tt uncurry} & {\tt uncurry} & {\tt uncurry} & {\tt uncurry} & {\tt uncurry} \\
 {\tt C}       & {\tt C}       & {\tt switch}  & {\tt C}       & {\tt swap\_arg} \\
 {\tt I}       & {\tt I}       & {\tt I}       & {\tt I}       & {\tt id\_fn} \\
 {\tt K}       & {\tt K}       & {\tt K}       & {\tt K}       & {\tt con\_fn} \\
 {\tt W}       & {\tt W}       & -             & {\tt W}       & {\tt dbl\_arg} \\
 {\tt (o)}     & {\tt (o)}     & {\tt (o)}     & {\tt (o)}     & {\tt ($<$*)} \\
 {\tt (\#\#)}  & {\tt (\#\#)}  & {\tt (**)}    & {\tt (F\_F)}  & {\tt pair\_apply} \\
 {\tt map}     & {\tt map}     & {\tt map}     & {\tt map}     & {\tt map} \\
 {\tt map2}    & {\tt map2}    & -	       & {\tt map2}    & {\tt bimap}	 \\
 {\tt \{funpow\}} & {\tt \{funpow\}} & {\tt fun\_pow} & {\tt \{funpow\}}  & {\tt funpow} \\
 {\tt itlist}  & {\tt itlist}  & {\tt fold}    & {\tt itlist}  & {\tt foldr} \\					
 {\tt rev\_itlist} & {\tt rev\_itlist} & {\tt revfold}  & {\tt rev\_itlist} & {\tt foldl} \\
 {\tt end\_itlist} & {\tt end\_itlist} & -     & {\tt end\_itlist} & {\tt foldr1} \\
 -             & -             & -	       & -             & {\tt foldl1}
\end{tabular}
\caption{Some of the functional programming library API components.}
\label{tab:fplib}
\end{table}

\subsubsection{Type, Term and Theorem Utilities}

Around 150 HOL type, term and theorem manipulation utilities are
provided (see Table~\ref{tab:utils} for a selection).

The bulk of these utilities are syntax functions for HOL types or
terms, for constructing, destructing and testing for a given syntactic
category.  Two levels of syntactic category are supported for both
types and terms.  Firstly, there are the primitive syntactic
categories, namely the type variables and type constant applications
for types, and variables, constants, function applications and lambda
abstractions for terms.  These are very widely used throughout the HOL
implementations.  Secondly, there are the basic syntactic categories
associated with the type constants and constants of predicate logic
and lambda calculus that feature in the theory platform.  Some of these
are also used heavily throughout the HOL implementations, but we
include support for all such syntactic categories in the API for
coherence with the theory platform and the API inference rules.

There are various ML bindings for HOL constants and base types featured
in the theory platform, and for commonly used HOL type variables.  Also
included are utilities for destructing a theorem into its assumptions
and conclusion parts, and for equality and alpha-equivalence comparison
of theorems.  There are also various type and term operations defined
that are essential for defining an inference kernel.  These include
calculating the type of a term, listing the type variables of a type,
testing for the alpha equivalence of two terms, and performing
variable and type variable instantiation.

The platform utilities for HOL terms are generally specified to work
modulo alpha equivalence in their arguments.  This was decided
because different systems generate bound variable names differently
when avoiding variable capture in type variable and variable
instantiation, and so this measure makes the API functions more robust
when ported.  An arbitrary bound variable name used in an operation in
one system could otherwise cause the equivalent operation in another
system to fail.   Note that hol90's {\tt free\_in}, which tests for
one term occurring free in another, does not work modulo alpha
equivalence, and so does not conform to the platform.

Note that there are various subtle differences between different
systems' utilities that can trip up casually ported code.  Examples
include ProofPower HOL's {\tt mk\_const} constructor, which does not
test that a constructed constant is well-formed, and hol90's and HOL4's
{\tt dest\_imp} and {\tt is\_imp}, which work for logical negation as
well as implication (although HOL4 has {\tt dest\_imp\_only} and
{\tt is\_imp\_only} for implication only).  The API chooses more
conventional behaviour.

\begin{table}
\footnotesize
\centering
\begin{tabular}{ c | c | c | c | c }
 hol90          & HOL4           & ProofPower     & HOL Light      & HOL Zero \\
 \hline
 {\tt type\_of} & {\tt type\_of} & {\tt type\_of} & {\tt type\_of} & {\tt type\_of} \\
 {\tt type\_vars\_in\_term}
                & {\tt type\_vars\_in\_term}
                                 & {\tt \{term\_tyvars\}}
                                                  & {\tt type\_vars\_in\_term}
                                                                   & {\tt term\_tyvars} \\
 {\tt aconv}    & {\tt aconv}    & {\tt ($\sim$=\$)}   & {\tt aconv}     & {\tt alpha\_eq} \\
 {\tt -}        & {\tt {rename\_bvar}} & -        & {\tt \{alpha\}} & {\tt rename\_bvar} \\
 {\tt free\_vars} & {\tt free\_vars} & {\tt frees} & {\tt frees}   & {\tt free\_vars} \\
 {\tt free\_varsl} & {\tt free\_varsl} & -        & {\tt freesl}   & {\tt list\_free\_vars} \\
 {\tt -}        & {\tt var\_occurs}
                                 & {\tt is\_free\_in}
				                  & {\tt \{vfree\_in\}} & {\tt var\_free\_in} \\	
 {\tt \{free\_in\}} & {\tt free\_in} & -          & {\tt free\_in} & {\tt term\_free\_in} \\
 {\tt all\_vars} & -             & -              & {\tt variables} & {\tt all\_vars} \\
 {\tt all\_varsl} & -            & -              & -              & {\tt list\_all\_vars} \\
 {\tt inst}     & {\tt \{inst\}} & {\tt \{inst\}} & {\tt \{inst\}} & {\tt tyvar\_inst} \\
 {\tt -}        & {\tt {rename\_bvar}} & -        & {\tt \{alpha\}} & {\tt rename\_bvar} \\
 -              & -              & {\tt \{var\_subst\}} & {\tt vsubst} & {\tt var\_inst} \\
 {\tt \{subst\}} & {\tt \{subst\}} & {\tt subst}  & {\tt subst}    & {\tt subst}
\end{tabular}
\caption{Some of the term utility API components.}
\label{tab:utils}
\end{table}

\subsubsection{Theory Extension and Listing Commands}

Around 40 theory extension and querying functions are provided.
This includes primitive theory extension commands for type
declaration, term declaration, constant definition, constant
specification and type constant definition.  On top of these, there
are a few basic derived theory extension commands, for example the
command to define a function constant using a universal quantifier
for the function arguments instead of a lambda abstraction.  Most
systems have more sophisticated extension commands, but these are
excluded from the platform because there is much variation in their
capability between systems.

Each system also provides querying commands to access information
about the theory extensions that have been made, although HOL Light
omits support for querying about primitive type constant definitions.
Such commands are essential for the approach for proof auditing
advocated in~\cite{audit}, and a complete set features in the API.

%

\subsubsection{Inference Rules}
\label{sec:apirules}

Around 100 basic inference rules are provided by the API (see
Table~\ref{tab:rules} for a selection).

It is sufficient for the platform inference rules to include just a
kernel of primitive rules\footnote{In the paper, we occasionally
abbreviate the term {\em inference rule} to {\em rule}.} that suffice,
when coupled with the axiom and definition theorems in the theory
platform, to implement the HOL deductive system.  Given our choice of
theory platform, any of the systems' primitive inference rules would
be sufficient.  However, efficiency is also a consideration.  If a
primitive rule of a given system were missing from the API, it would
have to be reimplemented in that system's outer platform in terms of
the API inference rules, and which would in turn need to be
implemented in terms of the system's primitives.  An execution of such
a recreated primitive could require 10 pre-platform rule applications
or more, resulting in an unacceptable performance penalty.  Thus we
choose to include the union of primitive rules from each system in the
platform (with the exception of one HOL Light primitive explained
below).  This principle qualifies around 35 rules for inclusion
in the platform.  Note that each system except HOL Zero has primitive
rules that are derive able in terms of other primitives, but are
included to improve the system's performance, which explains why the
union includes as many as 35.

Also included are around 15 other inference rules at roughly the
same level as the union of the primitive inference rules, including
the equality symmetry rule and the cut rule, for using the conclusion
of one theorem to eliminate an assumption in another.  A further 25
rules are included for performing equality congruence over certain
operators, in addition to the two that are present as a result of
being primitive inference rules.  For coherence, these fill out the
patchy provision in existing HOL systems with full coverage for the
HOL operators supported by the API syntax functions.

In addition, for natural arithmetic expressions there are conversions
provided for performing evaluation of operators applied to numeral
arguments for each of the 13 natural arithmetic operators featured in
the theory platform.  This is sufficient to provide complete coverage
of the primitive natural numeral arithmetic inference rules provided
by hol90 and ProofPower HOL (which represent numerals as constants).
This allows the platform to keep abstract the underlying
representation of numerals.

It is vital that the API specifies precise behaviour for each of
its inference rules.  There is a degree of variation in the behaviour
of various rules between systems. We outline here some ways in which
the platform promotes robustness in the details of the behaviour it
specifies for its inference rules.

As with the API's term utilities, its inference rules also work modulo
alpha equivalence, for the same reasons.  Note that the successful
execution of HOL Light's {\tt BETA} rule (not to be confused with its
{\tt BETA\_CONV} rule) can fail depending on the name used for a bound
variable in one of its arguments, and because of this it is excluded
from the API, despite being a primitive of HOL Light.  Fortunately, the
consequences on performance in HOL Light are minimal because {\tt BETA}
can be implemented purely in terms of {\tt BETA\_CONV}, which is in the
API.

It was also decided that API inference rules should not depend on the
presence of assumptions in their theorem arguments, also to help
robustness.  It is harmless for a rule to remove an assumption if it
can, and this should not result in failure in rules composed with it.
So, for example, the rule for discharging an assumption matching a
supplied term should not fail if the assumption is not present in the
theorem argument.  Note that ProofPower's classical contradiction rule
{\tt c\_contr\_rule} breaks this principle, but other systems'
equivalents do not.

There are also various other differences in behaviour between
seemingly equivalent rules in different HOL systems.  One particularly
extreme case is the rule for instantiating type variables, called
{\tt INST} in hol90, HOL4 and HOL Light, which is a primitive of every
HOL system.  In hol90, only type variables in the conclusion are
instantiated.  In HOL Light and HOL4, non-variable types in the
instantiation list argument do not cause failure.  And in ProofPower
HOL, any free variables that would otherwise become equal as a result
of the instantiation are renamed.  None of these idiosyncrasies exist
in the API version.

\begin{table}
\footnotesize
\centering
\begin{tabular}{ c | c | c | c | c }
 hol90             & HOL4              & ProofPower        & HOL Light         & HOL Zero  \\
 \hline
 {\tt ASSUME\star} & {\tt ASSUME\star} & {\tt asm\_rule\star}
                                                 & {\tt ASSUME\star} & {\tt assume\_rule\star} \\	
 {\tt BETA\_CONV\star} & {\tt BETA\_CONV\star}
                                       & {\tt simple\_$\beta$\_conv\star}
                                                           & {\tt BETA\_CONV}  & {\tt beta\_conv\star} \\
 {\tt CCONTR\star} & {\tt CCONTR\star} & {\tt \{c\_contr\_rule\}}
                                                           & {\tt CCONTR}      & {\tt ccontr\_rule} \\
 {\tt CHOOSE\star} & {\tt CHOOSE\star} & {\tt simple\_$\exists$\_elim}
                                                           & {\tt CHOOSE}      & {\tt choose\_rule} \\
 {\tt CONJ\star}   & {\tt CONJ\star}   & {\tt $\wedge$\_intro}
                                                           & {\tt CONJ}        & {\tt conj\_rule} \\
 {\tt CONJUNCT1\star} & {\tt CONJUNCT1\star}
                                       & {\tt $\wedge$\_left\_elim}
                                                           & {\tt CONJUNCT1}   & {\tt conjunct1\_rule} \\
 {\tt CONJUNCT2\star} & {\tt CONJUNCT2\star}
                                       & {\tt $\wedge$\_right\_elim}
                                                           & {\tt CONJUNCT2}   & {\tt conjunct2\_rule} \\
 {\tt CONTR\star}  & {\tt CONTR}       & {\tt contr\_rule} & {\tt CONTR}       & {\tt contr\_rule} \\
 -                 & -                 & -                 & {\tt DEDUCT\_ANTISYM\_RULE\star}
                                                                               & {\tt deduct\_anitsym\_rule} \\
 {\tt DISCH\star}  & {\tt DISCH\star}  & {\tt $\Rightarrow$\_intro\star}
                                                           & {\tt DISCH}       & {\tt disch\_rule\star} \\
 {\tt DISJ1\star}  & {\tt DISJ1\star}  & {\tt $\vee$\_right\_intro} & {\tt DISJ1}  & {\tt disj1\_rule} \\
 {\tt DISJ2\star}  & {\tt DISJ2\star}  & {\tt $\vee$\_left\_intro}  & {\tt DISJ2}  & {\tt disj2\_rule} \\
 {\tt DISJ\_CASES\star} & {\tt DISJ\_CASES\star}
                                       & {\tt $\vee$\_elim} & {\tt DISJ\_CASES} & {\tt disj\_cases\_rule}
\end{tabular}
\caption{Some of the inference rule API components.  Primitive rules in a
         given system are marked with~{\tt \footnotesize \star}.}
\label{tab:rules}
\end{table}

\subsubsection{Parsing and Pretty Printing}

Around 20 functions supporting parsing and pretty printing are provided
in the API.  This includes functions for parsing strings into HOL types
and terms, and printers for types, terms and theorems.  There is also
support for setting the fixity of HOL functions and type operators.
The fixities supported exceed what is provided by hol90, ProofPower HOL
and HOL Light, but do not extend to the full range of fixities supported
by HOL4.  There are plans to extend the platform to support all of HOL4's
fixities.

%

\section{Implementation}

\subsection{Architecture}
\label{sec:arch}

For a legacy system to conform to an API, its source code must be
adapted so that every component of the API is implemented in the
system.  For the Common HOL API, we use a software architecture for
adapting legacy HOL systems that is designed with the three goals
of minimising implementation effort, enabling API-level
virtualisation, and facilitating the demonstration that the adapted
system exhibits precisely the same behaviour as the legacy system.
 
To achieve this, we choose an appropriate point in the build of
the legacy system that corresponds to the level of the API (the
{\em platform level}), and insert an ML module for the API components
(the {\em platform module}) at this point.  All legacy source code
occurs either before or after the platform level (respectively called
the {\em pre-platform} and {\em post-platform} code) and stays exactly
the same.  Keeping the pre- and post-platform code the same makes it
easier to argue that the system's behaviour has not been altered.

In the platform module, we define the API in terms of pre-platform
functionality.  Any API components not precisely implemented as a
pre-platform component must be implemented here.  This includes
components missing from the legacy system, or with imprecisely
corresponding equivalents in the pre-platform code or that are
implemented as post-platform code.  For any implemented as
post-platform code, the full tree of post-platform code used to define
it can be shifted into the platform module, or, if this is too big,
then a more succinct version can be implemented specially for the
platform.  The code for post-platform API components can then be
deleted from its original position in the source code (thus the
post-platform code remains the same except for deleted code that
occurs in the platform module).

In our architecture, all post-platform code implementing primary
functionality is implemented in terms of the API.  This enables the
API to act as a virtualisation layer through which all primary
functionality is executed.  This virtualisation layer can then be
used for recording proofs as they are executed, before exporting them
to proof objects.  In order to achieve this and keep the
post-platform code the same, we must somehow have a way of referring
to pre-platform code that is used by post-platform code but is not in
the API.  We do this by implementing a module immediately after the
platform module in the build that re-implements all such pre-platform
code in terms of the platform, overwriting the pre-platform code.
We call this the {\em outer platform} module.

In arguing that the system's behaviour has not altered in the
API-adjusted version of the system, we must justify why any
reimplementation of post-platform code in the platform module, and
any reimplementation of pre-platform code in the outer platform
module, preserves functionality.

Given that the API components correspond to classic basic components
of a HOL system that tend to be implemented towards the start of the
build of the system, finding an appropriate insertion point for the
platform level tends to be fairly straightforward.  It is to be found
after the definition of the HOL type and term datatypes and basic
utilities for manipulating them, the inference kernel, the initial
theory and the parser and pretty printer.  It is typically before
the derived inference rules for predicate logic and the theory for
pairs and natural numbers, which would need to be moved to
or recreated in the platform module.

%

\subsection{Adapting HOL Light}
\label{sec:hladapt}

We now describe how we adapted HOL Light SVN release 197 to conform to
the platform.  The reader may find it instructive to download the
adapted system~\cite{hollightplus}.

The platform level in the HOL Light build file was chosen between the
source files {\tt parser.ml} and {\tt equal.ml}.  About 1,000 lines of
post-platform code implementing platform components were moved into the
platform module.  Much of this was derived inference rules implemented
using lemmas proved using HOL Light's automated proof facilities.
Instead of recreating these facilities inside the platform module, we
employed Common HOL proof porting to export the proofs of these lemmas
as proof objects, which were then hand-translated into a total of
around 400 lines of forwards style proof script in the platform module.
An alternative approach was used to recreate the 13 evaluation rules
for natural numeral arithmetic, whose implementation in
{\tt calc\_num.ml} involves lemmas proved in hundreds of lines of proof
script.  Instead of exporting proof objects for these lemmas, the
inference rules were given a completely different implementation in the
platform module, ported from HOL Zero in about 800 lines.

About 1,000 lines of code were required to fill out platform components
missing from HOL Light.  For those components with an approximate
equivalent already in HOL Light, the existing component was used in the
implementation of the platform variant (e.g.\ see Figure~\ref{fig:insttype}),
to ensure that the platform variant had roughly the same performance as
the original.  Those components with no approximate HOL Light version
were ported from HOL Zero.  In total, the components ported from HOL Zero
required about 1,350 lines of supporting source code to be ported from
HOL Zero, mainly involving forwards proof to prove lemmas.  The platform
module interface is written in about 500 lines of code.

For the outer platform, primitive inference rules and theory commands
that do not correspond to platform components must be precisely recreated
in terms of the platform.  In HOL Light, this involves the
{\tt INST\_TYPE} and {\tt BETA} rules and all the theory commands.
Also, non-platform theorems used to define platform theory needed to
be recreated.  In total, the outer platform required around 800 lines of code.

Overall, the platform and outer platform modules involved around 6,000
lines of source code, including the platform module interface.  This
took around two weeks of effort to create.  The code was mostly
systematically produced, being either moved from other parts of HOL
Light, ported from HOL Zero, translated from proof object files, or
simply a listing of platform components.  The only code requiring
creative thought was in the platform module variants of components with
approximate equivalents already in HOL Light, and in much of the outer
platform, totalling to around 1,000 lines.

\begin{figure}
\footnotesize
\begin{verbatim}
  let INST_TYPE1 theta th =
    let () = if (forall (is_vartype o snd) theta)
               then failwith "INST_TYPE: Non-type-variable in instantiation domain" in
    INST_TYPE theta th;;
\end{verbatim}
\caption{Using HOL Light's original {\tt INST\_TYPE} in the definition of the
         platform variant.}
\label{fig:insttype}
\end{figure}

\section{Use Cases}

In this section, we report on two use cases for the Common HOL Platform
in assisting manual ports of source code between platform-adapted
HOL systems.  In both cases, the port was from HOL Light to HOL Zero.
This is on the easy end of the difficulty spectrum in inter-HOL-system
code porting, because both systems are implemented in the same dialect
of ML, i.e.\ OCaml, and because the target system, HOL Zero, is almost
a blank canvas with very little post-platform code to consider. Other
HOL systems have considerable post-platform code, and porting should
attempt to reuse any pre-existing code if it is straightforward to do
so, to avoid creating an almost duplicate stack of supporting
functionality in the target system.  However, both ports described here
would still be difficult without the support of the platform, and so
the use cases provide useful insight.

\subsection{Legacy Code Port: HOL Light Rewriting Mechanism to HOL Zero}

\begin{figure}[!t]
\footnotesize
\begin{verbatim}
  let mk_rewrites =
    let IMP_CONJ_CONV = REWR_CONV(ITAUT `p ==> q ==> r <=> p /\ q ==> r`)
    and IMP_EXISTS_RULE =
      let cnv = REWR_CONV(ITAUT `(!x. P x ==> Q) <=> (?x. P x) ==> Q`) in
      fun v th -> CONV_RULE cnv (GEN v th) in
    let collect_condition oldhyps th =
      let conds = subtract (hyp th) oldhyps in
      if conds = [] then th else
      let jth = itlist DISCH conds th in
      let kth = CONV_RULE (REPEATC IMP_CONJ_CONV) jth in
      let cond,eqn = dest_imp(concl kth) in
      let fvs = subtract (subtract (frees cond) (frees eqn)) (freesl oldhyps) in
      itlist IMP_EXISTS_RULE fvs kth in
    let rec split_rewrites oldhyps cf th sofar =
      let tm = concl th in
      if is_forall tm then
        split_rewrites oldhyps cf (SPEC_ALL th) sofar
      else if is_conj tm then
        split_rewrites oldhyps cf (CONJUNCT1 th)
          (split_rewrites oldhyps cf (CONJUNCT2 th) sofar)
      else if is_imp tm & cf then
        split_rewrites oldhyps cf (UNDISCH th) sofar
      else if is_eq tm then
        (if cf then collect_condition oldhyps th else th)::sofar
      else if is_neg tm then
        let ths = split_rewrites oldhyps cf (EQF_INTRO th) sofar in
        if is_eq (rand tm)
        then split_rewrites oldhyps cf (EQF_INTRO (GSYM th)) ths
        else ths
      else
        split_rewrites oldhyps cf (EQT_INTRO th) sofar in
    fun cf th sofar -> split_rewrites (hyp th) cf th sofar;;
\end{verbatim}
\caption{A sample of legacy source code from HOL Light's {\tt simp.ml}.}
\label{fig:sample1}
\end{figure}

\begin{figure}[!t]
\footnotesize
\begin{verbatim}
  let mk_rewrites =
    let imp_conj_conv = rewr_conv imp_imp_thm
    and imp_exists_rule =
      let cnv = rewr_conv imp_exists_rule_thm in
      fun v th -> conv_rule cnv (gen_rule v th) in
    let collect_condition oldhyps th =
      let conds = subtract (asms th) oldhyps in
      if conds = [] then th else
      let jth = foldr disch_rule conds th in
      let kth = conv_rule (repeatc imp_conj_conv) jth in
      let cond,eqn = dest_imp(concl kth) in
      let fvs = subtract (subtract (free_vars cond) (free_vars eqn))
                         (list_free_vars oldhyps) in
      foldr imp_exists_rule fvs kth in
    let rec split_rewrites oldhyps cf th sofar =
      let tm = concl th in
      if is_forall tm then
        split_rewrites oldhyps cf (spec_all_rule th) sofar
      else if is_conj tm then
        split_rewrites oldhyps cf (conjunct1_rule th)
          (split_rewrites oldhyps cf (conjunct2_rule th) sofar)
      else if is_imp tm & cf then
        split_rewrites oldhyps cf (undisch_rule th) sofar
      else if is_eq tm then
        (if cf then collect_condition oldhyps th else th)::sofar
      else if is_not tm then
        let ths = split_rewrites oldhyps cf (eqf_intro_rule th) sofar in
        if is_eq (rand tm)
        then split_rewrites oldhyps cf (eqf_intro_rule (gsym_rule th)) ths
        else ths
      else
        split_rewrites oldhyps cf (eqt_intro_rule th) sofar in
    fun cf th sofar -> split_rewrites (asms th) cf th sofar;;
\end{verbatim}
\caption{The translation into HOL Zero of the legacy code sample from
         {\tt simp.ml}.}
\label{fig:sample2}
\end{figure}

In our first use case, we ported HOL Light's entire rewriting
apparatus to HOL Zero.  This is defined relatively early on in HOL
Light's post-platform code, but provides vital functionality that is
used throughout the rest of the system, and goes far beyond what HOL
Zero is capable of in terms of proof automation.  It is implemented
in 360 lines of code, in the HOL Light source file {\tt simp.ml}, and
relies on 60 lines of code defining discrimination nets, and a
further 300 lines of post-platform code defining supporting
functionality such as conversion combinators.  Thus there was a total
of 720 lines to port, but this would probably be less if porting to
another HOL system because it would already support conversion
combinators.  See Figures~\ref{fig:sample1} and \ref{fig:sample2} for
a sample of 32 lines from the port.

The manual port was carried out in about 2 hours 30 minutes of
effort.  Note that this time does not include approximately 30
minutes of effort required to extract out the 360 lines of HOL Light
supporting code prior to the port.  The porting itself involved
systematically looking up HOL Zero equivalents of HOL Light platform
functions, and renaming accordingly.  HOL Light's uppercase names,
that don't conform to normal OCaml lexical syntax, also needed to be
converted to lowercase names.  Instantiation lists, which have
old-to-new ordering in HOL Zero but new-for-old ordering in HOL Light,
needed to be switched around.  The datatype constructors for types
and terms, which are visible outside their defining module in HOL
Light but not in HOL Zero, required some pattern matches to be
replaced with abstract destructors and if-expressions.  The
function {\tt term\_match} name-clashed with a pre-existing HOL Zero
function, and so was renamed to {\tt hl\_term\_match}.

HOL Light non-conformant versions of platform functions, such as its
{\tt variant} function, required special attention.  Unlike the
platform equivalent, this function does not fail if its avoidance list
contains non-variables, and so the code was adapted to either filter
them out or check that non-variables are not possible from program
context.  Other complications included two uses of HOL Light's
intuitionistic tautology prover, {\tt ITAUT}.  It was decided to keep
this function outside the scope of the port, despite it being used to
prove two lemmas, to reduce the amount of supporting code.  For the HOL
Zero version, one of the lemmas already existed in HOL Zero's small
library of predicate logic theorems, and the other was proved in 10
minutes in a 16-line proof using HOL Zero's forward inference rules.

After the port was completed, it was tested on various rewriting
examples, and one error was found.  This took 45 minutes of debugging
to track down and correct, and was due to a quirk in the failure
exception returned by HOL Light's {\tt rev\_assoc} function, which
has error message text {\tt "find"} (instead of {\tt "rev\_assoc"}).
This particular error message was explicitly trapped in the HOL Light
code, but naively porting this to HOL Zero didn't work because its
equivalent function, {\tt inv\_assoc}, uses error message text
{\tt "inv\_assoc"}.  As explained in Section~\ref{sec:lib}, this
aspect of porting is not catered for by the platform, and must be
done manually.


\subsection{New Code Port: HOL Light Proof Importer to HOL Zero}

In the second use case, we used the platform to port HOL Light's
importer for Common HOL proof objects.  This was a fundamentally
easier exercise because the proof importer is written specifically
in terms of the API, and because Common HOL proof porting works at
the level of platform inference rules itself.  The proof importer is
implemented in 2,200 lines of code.

It took about 1 hour 15 minutes to perform the porting.
Despite the source code being three times longer than in the legacy code
port, it took only half the time.  The easier nature of the task
meant that everything went smoothly first time.  The effort consisted
almost entirely of systematically applying search-and-replace to
replace HOL Light platform function names with HOL Zero equivalents
and carrying out manual adjustments for functions that take their
arguments differently in the different systems.

The resulting source code was tested by importing into HOL Zero the
text formalisation part of the Flyspeck project, as part of a partial
audit of the project as described in~\cite{audit}.  This involved the
tens of millions of platform-level inference rule steps.  The import
into HOL Zero worked first time, suggesting the code was ported
correctly.

%

\section{Conclusions}

In defining a standard for basic theory and programming components, the
Common HOL Platform is attempting to lay the foundation for much better
portability between HOL systems, both in terms of porting proofs and
porting source code.  The feasibility of large scale proof porting has
already been established by others, but arguably there is scope for
doing better still, given a better foundation.  However, the
feasibility of quick and reliable source code porting has not been
explored until now.

In this paper, we have given an overview of the platform's components
and explained the reasons behind some of the careful design decisions
made.  We have also demonstrated using the platform in two use cases of
manually porting source code from HOL Light to HOL Zero, one for legacy
code and one for new code written specially for the platform.  In both
cases, several hundred lines of code were successfully and reliably
ported within a few hours.  Much of the effort normally involved in a
manual port is removed, because almost all that needs to be considered
is functionality implemented above the platform level.  Finding
corresponding low-level components in the two systems, and the subtle
ways in which they can differ, has already been taken care of by the
platform.  As far as we are aware, this represents a leap in the
productivity of source code porting between HOL systems, even when
accounting for it being less challenging than the general porting case
due to both systems being implemented in the same dialect of ML and due
to HOL Zero effectively being a blank canvas.

It would be interesting to see how far HOL source code porting could be
pushed.  Certainly it is feasible to port more challenging parts of HOL
Light to HOL Zero.  Obvious candidates are the subgoal package, the
intuitionistic tautology checker and the powerful {\tt MESON\_TAC}.
Implementing the latest version of the platform for hol90,
HOL4 and ProofPower HOL, and porting to these systems is another
challenge worth pursuing.  The platform has already been designed with
these systems in mind, and it would at least enable Common HOL proof
exporters and importers to be quickly ported to these systems.

One insight that comes from looking at code from the various HOL
systems is how much the subgoal package is used in the implementation
of other parts of HOL systems, suggesting that it should be part of the
API.  This should be a fairly easy extension to make, since beyond the
implementation of an initial few tactics, code using it appears to
operate at the abstract level using tacticals, rather than use the
inner workings that differ between HOL systems.  Another change worth
making is to update the platform for the reform to primitive theory
extension currently underway in various HOL systems~\cite{rob}.  And
finally, catering for Isabelle/HOL must be a long term priority.  This
would probably require a significant overhaul of the platform to fit
with such a different system, but if done well it would pay dividends
to have good portability between the widest used HOL system and the
rest of the family.

The systematic manner in which the porting can be carried out lends
itself to automation, or at least to partial automation.  The most
difficult to automate is probably the intelligent use of the target
system's legacy supporting code to avoid the ugly situation of
creating two parallel stacks of code implementing effectively the same
thing.  Thus partial automation looks a more realistic prospect.  We
believe there are no fundamental difficulties in automatically porting
between ML dialects, because the subsets of ML that tend to be used in
the implementation of HOL systems are trivially corresponding between
OCaml and SML.  So we see there being good prospects for reducing
further the time taken to reliably port source code, even in more
challenging cases.

\end{document}